\documentclass[12pt,a4]{article}
\usepackage{epsfig}
\usepackage{rotating}

\begin{document}
\title{Cosmological Models with Shear and Rotation}
\author{SHWETABH SINGH$^{*}$}
\date{}
\maketitle
{\vskip 0.2in}
\centerline{Physics Department}
\centerline{Indian Institute of Science, Bangalore-560012}
{\vskip 4.65in}
\centerline{$^{*}$The work was carried out when the author was a student at}
\centerline{Physics Department, Panjab University, Chandigarh-160014}
\newpage

\centerline{\bf KEYWORDS}
\medskip
\noindent Cosmological Models, Singularity, Rotation, Centrifugal Force, Shear
{\vskip 0.7in}

\centerline{\bf ABSTRACT}
\medskip

Cosmological models involving shear and rotation are considered, first in the General Relativistic and then in the Newtonian framework with the aim of investigating singularities in them by using numerical and analytical techniques. The dynamics of these rotating models are studied. It is shown that singularities are unavoidable in such models and that the centrifugal force arising due to rotation can never overcome the gravitational and shearing force over a length of time.

{\vskip 0.7in}
\centerline{\bf I. INTRODUCTION}
\medskip

Cosmological models with rotation have drawn attention in the past for various reasons. G$\ddot{o}$del first proposed a rotating universe model, with the aim of demonstrating that general relativity does not incorporate Mach's principle.Another reason for proposing the rotating models has been with the view to avoid a singularity in the universe.This approach was pioneered by Heckmann and Sch$\ddot{u}$cking. The basis for this line of thought has been that the centrifugal force arising due to rotation would prevent a collapse of the universe. Finally, since all astrophysical systems have rotation in them so it has been cojectured that there might be rotation in the universe too.

The present work will deal with the investigation of singularities in a special class of metrics namely the Heckmann Sch$\ddot{u}$cking metrics in the Relativistic and Newtonian framework.Although in general relativity a singularity is unavoidable, as shown by Hawking and Penrose, the study of the dynamics of rotating models does yield an idea of how far rotation may be able to prevent a collapse. The centrifugal force may prevent collapse along the axis but the collapse may nonetheless take place perpendicular to the axis of rotation. The first part of the work deals with the relativistic Heckmann Sch$\ddot{u}$cking model while in the second part the Newtonian analogues of relativistic models will be looked at.
  
{\vskip 0.7in}
\centerline{\bf II. THE RELATIVISTIC HECKMANN SCH$\ddot{U}$CKING METRIC}
\medskip

The most common approach to describe the cosmology of the universe is  on the basis of the assumption of homogenity and isotropy of the universe which leads to the Robertson - Walker metric

\begin{equation}
         ds^2=dt^2-a^2(t)[dr^2/(1-kr^2) + r^2(d\theta^2 +sin\theta^2d\phi^2)],
\end{equation}
\noindent where $k$=1,0,-1.

\noindent This line element when coupled with the field equations, 
\begin{equation}
    R^{\mu\nu}-\frac{1}{2}g^{\mu\nu}R = -\kappa T^{\mu\nu}-\Lambda g^{\mu\nu},
\end{equation}
\noindent can give the value of $a(t)$ where $T^{\mu\nu}$ is given by 
\begin{equation}
    T^{\mu\nu}= (p +\rho + \frac{4}{3}u)\frac{dx^{\mu}}{dt}\frac{dx^{\nu}}{dt} 
- (p+\frac{u}{3})g^{\mu\nu},
\end{equation}

\noindent where $\rho$ and $u$ are the matter and radiation densities and $p$ is the pressure.

This form of the solution gives rise to a universe with a singular origin. Can the singularity be averted if  one considers a modification in the assumptions 
governing the choice of the line element? The postulate regarding isotropy 
of the universe is replaced by anisotropy which may be due to rotation and shear. The world lines of the galaxies are still geodesics along which the world time is measured but the t=constant surfaces are no longer orthogonal to the geodesics. This gives rise to cross product terms of the type $g_{\mu4}dx^{\mu}dt (\mu= 1,2,3)$ in the metric.

One such type of  metric was proposed by G$\ddot{o}$del (1949),
\begin{equation}
   ds^2=dt^2+2e^{x_1}dtdx^2 - (dx^1)^2 +\frac{1}{2}e^{2x_1}(dx^2)^2 -(dx^3)^2.
\end{equation}

The solution of the field equations with this metric indicate a rotating and 
stationary universe with the angular velocity of the various components given 
by
\begin{equation}
   \omega^{\mu} = 
\frac{\epsilon^{\mu\nu\lambda\kappa}}{6\sqrt{g}}a_{\nu\lambda\kappa}.
\end{equation}

\noindent The non-vanishing of $\omega^3$ shows a rotation about the $x^3$= constant 
axis.

The failure of this model to account for the red shift of galaxies because of 
its stationary character prompted Heckmann and Sch$\ddot{u}$cking (1958) to give a 
generalisation of the G$\ddot{o}$del's metric to obtain a universe with a non-singular 
origin and also to account for the red shift by making this model non-static. They hoped to find finite oscillating universes with  maximum and minimum radii.

One such model having G$\ddot{o}$del's model as a special case, has the line 
element 

\begin{equation}
ds^2 = dt^2 + 2 e^{x^1} dt {dx}^2 - c_{11} (t) (dx^1)^2 - 2 c_{12} (t) e^{x^1} 
dx^1 dx^2 + \alpha c_{11} (t) e^{2{x_1}} (dx^2)^2 - S^2 (t) (dx^3)^2.
\end{equation}

\noindent which when solved with the field equations gives the following set of 
differential equations :
\begin{equation}
\frac{1}{4R^2} \left[ ( \dot{c_{12}}+1)^2 + \alpha\dot{c_{11}}^2 -4\alpha c_{11} \right] 
+ \frac{\dot{S}}{S} ( \frac{ 2{c_{12}}+\dot{c_{11}}}{2R^2} - \frac{\dot{R}}{R} 
) = - \Lambda - \frac{\alpha^2}{RS},
\end{equation}

\begin{equation}
\dot{c_{12}}c_{11} -\dot{c_{11}}c_{12} -c_{11}=  -\frac{\alpha}{RS},
\end{equation}

\begin{equation}
\frac{\ddot{R}}{R} +\frac{\ddot{S}}{S} 
-\frac{1-\dot{c_{12}}^2-\alpha\dot{c_{11}}^2}{2R^2}= \Lambda 
-\frac{\alpha^2}{2RS},
\end{equation}
where
\begin{equation}
R^2= c_{11} -\alpha c_{11}^2 -c_{12}^2,
\end{equation}
\begin{equation}
\rho= \frac{\alpha^2}{RS},
\end{equation}
\begin{equation}
\omega= \frac{1}{\sqrt{2}R}.
\end{equation}

Heckmann and Sch$\ddot{u}$cking did not solve these equations which are too complicated for an analytical solution. Narlikar (1960) showed, however, that these models would not fulfill their original purpose. Later the interest in these models waned as the singularity theorems gained wide acceptance. Nevertheless we return to the Heckmann Sch$\ddot{u}$cking model to investigate its dynamical behaviour by numerical methods, to see how the singularity actually develops.

The above equations were thus solved numerically with the initial values taken in 
correspondence with the G$\ddot{o}$del (1949) metric except for the introduction of a 
$\dot{c_{11}}$term to introduce a non static character in the metric.Thus,
\begin{equation}
c_{11}(0)=1,
\end{equation}
\begin{equation}
c_{12}(0)=0,
\end{equation}
\begin{equation}
\Lambda=-\frac{1}{2},
\end{equation}
and different values of $\dot{c_{11}}$ and $\alpha$ were taken to highlight 
different cases. The parameters were varied as follows:
\begin{equation}
\dot{c_{11}}=0.1,0.01,0.001
\end{equation}
\begin{equation}
\alpha =\frac{1}{2}, -\frac{1}{2}
\end{equation}
in correspondence with the two seperate cases pointed out by Narlikar (1960).

The results for $\alpha=\frac{1}{2}$ as shown in figure (1) shows a collapse in $R$ and figure (2) reveals the small oscillations $R$ makes before the collapse. Figure (3) shows a continuously shrinking $x^3$ dimension.
The coefficient of $(dx^3)^2$ in the metric goes to zero linearly showing a one 
dimesional singularity. $R$, which shows the extent of the universe in the other 
two dimensions also goes to zero[Figure (1)] pointing to a singularity in the volume after 
a period of oscillations since $V\propto R^{2}S$. The equations also imply a 
rapidly increasing angular velocity by consequence of $\omega= 
\frac{1}{\sqrt{2}R}$ as pointed out by G$\ddot{o}$del (1950) in one of his discussions of the 
properties of the metric.

The oscillations are sustained for a longer period of time if the initial rate 
of change of $c_{11}$with respect to time is large as this delays $R$ from 
falling to zero as a consequence of 
\begin{equation}
R^2= c_{11} - \alpha c_{11}^2 - c_{12}^2,
\end{equation}
becoming smaller and smaller more slowly.
\par
It can also be seen from figures (4) and (5) that for the case $\alpha=-\frac{1}{2}$ oscillations are not 
possible in the G$\ddot{o}$del like universe and for the same other parameters the 
universe shrinks to a two dimensional singularity with a rapidly increasing $S$[Figure (5)], 
the coefficient of $(dx^3)^2$ and a decreasing $R$[Figure (4)].

The volume of the universe given by $V\propto R^{2}S$ also goes to zero 
suggesting that a singularity results in this case too.The results for 
$\alpha=-\frac{1}{2}$ are in agreement with the analytical argument of 
Narlikar (1960) where he has shown that for $\alpha =-\frac{1}{2}$ the universe does not oscillate between finite limits by considering $ (7), (8)$, and $(9)$ coupled with the signature conditions. This form of a universe leads to a situation similar to the isotropic case.
\par
Another result which may be directly seen from the numerical simulation is that 
the universe for $\alpha=-\frac{1}{2}$ is inherently unstable and collapses 
about an order of magnitude faster than for $\alpha=\frac{1}{2}$, all other 
parameters being comparable.

{\vskip 0.7in}
\centerline{\bf III. NEWTONIAN COSMOLOGIES WITH ROTATION}
\medskip

One of the first works done on developing cosmology on Newtonian terms was by Milne and McCrea (1934). Their work was based on the concept of isotorpy and homogenity in the universe and they showed that models analogous to the Friedmann Robertson Walker model could be obtained by using the Newtonian theory too. Heckmann and Sch$\ddot{u}$cking (1955) formulated the Newtonian equations of cosmology by solving the following three equations used to describe the general behaviour of a homogenous universe, but doing away with the assumption of isotropy:

\begin{equation}
\frac{\dot{\rho}}{\rho} + div.{\bf v} = 0,     
\end{equation}

\begin{equation}
{\bf v} + {(\bf v}.{\bf \nabla)} {\bf v} = -\phi,   
\end{equation}

\begin{equation}
\nabla^2 \phi + \lambda = 4 \pi \rho G.
\end{equation}

The Newtonian equations of cosmology formulated by them include both  shear and  rotation terms. Thus while rotation prevents the universe from collapsing, shear has the opposite effect. This is analogous to the general relativistic result derived by Raychaudhuri (1955). They tried to avoid singularity by setting the shear equal to zero which is possible in the Newtonian framework. They were able to show that it is possible to avoid singularities because the rotation term dominates as $R\rightarrow 0$ , hence preventing a collapse.

This freedom to take the shear equal to zero in Newtonian models does not exist in General Relativistic models as was shown by Ellis (1967). Hence, it is not possible to avoid singularities in the General Relativistic models which give a more accurate description than their Newtonian analogs, in accordance with the Hawking-Penrose theorem (1975).

However in 1963, Narlikar formulated the gravitational force approach which was based on the inverse square law rather than the Poisson equation,
\begin{equation}
\nabla^2 \phi = 4 \pi G \rho.
\end{equation}
\noindent This puts a greater restriction on $\phi$ and implies that even though it might be possible to put the shear equal to zero initially, the time dependence of the shear has the effect of it acquiring some finite value in the course of time. Thus it is not obvious that singularity can be prevented.

By the axioms of Newtonian mechanics we have a Euclidean space with rectangular co-ordinates $x_{\mu} (\mu =1,2,3) $ and an even-flowing uniform time, $t$. Thus at any time $t$ the universe will present the same large scale view to all observers whose motion is idealized as the streaming of an ideal fluid. The velocity-distance relation can be written in the form,
\begin{equation}
v_{\mu} = H_{\mu\nu} x_{\nu},
\end{equation}

\noindent where $H_{\mu\nu}$ is a function of $t$ only, which appears in analogy with the Hubble's constant in the Robertson-Walker cosmology, but is direction dependent in accordance with the assumption of anisotropy.
After writing the solution of (23) in the form,
\begin{equation}
x_{\mu} = a_{\mu\nu}(t) x^{0}_{\nu},
\end{equation}
$\Delta = det||a_{\mu\nu}||$ and solving (19) and (20), Narlikar (1963) arrived at the following equation of motion,
\begin{equation}
\Delta \frac{\ddot{a_{\mu\nu}}}{2} = -\frac{4 \pi G \rho_{0} a_{\mu\nu}}{3}
\end{equation}

\noindent After replacing the time $t$ by a dimesionless co-ordinate,
\begin{equation}
\tau = (\frac{4 \pi G \rho_{0}}{3})^{\frac{1}{2}} t
\end{equation}

\noindent and continously differentiating (25), Narlikar (1963) was able to reduce the six equations of motion to a single fourth order equation:
\begin{equation}
\Delta^2 \Delta^{\prime \prime \prime \prime} + 7 \Delta \Delta^{\prime \prime} - 4 \Delta^{\prime 2} + 9 \Delta = 0.
\end{equation}

\noindent This fourth order nonlinear differential equation may be reduced by the following transformations,
\begin{equation}
\Delta = F^2,    (\frac{dF}{d\tau})^2= X(F),   F=e^{U},  \frac{dX}{dU}= Y(X),
\end{equation}

\begin{equation}
X = F'^2 ,    Y= 2 F F^{\prime \prime},
\end{equation}

\noindent to the following second order equation,
\begin{equation}
XY^2(\frac{d^2Y}{dX^2}) + XY(\frac{dY}{dX})^2 + (X + \frac{Y}{2})Y\frac{dY}{dX} +Y^2 - 2XY + 7Y - 2X + 9 = 0.
\end{equation}

The above equation was solved numerically by Narlikar (1963) for various initial conditions and all of them did show a singularity. Davidson and Evans (1973) investigated (25) further and all their numerical and analytical results also show a singularity.

But, since numerical analysis cannot be exhaustive of all initial conditions, (30) was analysed asymptotically to arrive at a more general result. We take the form of the asymptote as,
\begin{equation}
Y = mX + K
\end{equation}

Substituting $Y$ in (30) and equating the coefficients of $X^2$ and $X$ to zero we find the following asymptotes to the curves
\begin{equation}
Y = K_{1},
\end{equation}
\begin{equation}
Y = -2X + K_{2},
\end{equation}
\begin{equation}
Y = \frac{2}{3} X - 3.
\end{equation}

Here $K_{1}$ and $K_{2}$ are constants and (34) is an exact solution and was also found by Narlikar (1963). The curves with asymptotes $Y=-2X +K_{1}$ were also investigated by Narlikar (1963), and were found to go asymptotically as $Y=-2X$ as is shown in  figure (6). However the curves with asymptotes $Y= K_{1}$ were missed in that work.

Substituting the value of Y from (29) in (32), we get,
\begin{equation}
F F^{\prime \prime} = k.
\end{equation}

Solving the above equation by integrating leads to,
\begin{equation}
\frac{dF}{\sqrt{ln F}} = c_{1} d \tau.
\end{equation}

Substituting $x^{2}=ln F$, we can get the equation in the form of the integral of the error function,

\begin{equation}
\int e^{x^{2}}dx = c_{2} \tau,
\end{equation}

\noindent where $c_{2}$ is some constant independent of $\Delta$ and $\tau$.

It may be seen from (37) that  $\tau$ has a range of values from $-\infty$ to $\infty$ being the time co-ordinate. By readjusting  the time scale we can always arrange that  at some point in the range of $\tau$, $\Delta$ must become equal to zero which will imply a singularity.

The two other asymptotes of the curve have already been investigated numerically by Narlikar (1963) and Davidson and Evans (1973) and were shown to be singular. The asymptote $Y$= constant does indeed correspond to a few numerical solutions which were compiled by Davidson and Evans (1973).

Another way of looking at the singular nature of the solution is by investigating the case when asymptotically $Y=-2X$. Then we can take,
\begin{equation}
Y= -2X + c = -2 X Z(X)
\end{equation} 
where the function $Z(X)\rightarrow 1$ as $X\rightarrow \infty$. Thus we get,
\begin{equation}
Z(X) = 1 - \frac{c}{2X}
\end{equation}
where $c$ is a constant independent of $X$.

Substituting these values of $Y$ and its first and second derivatives in (30) we see that the quadratic terms of (30) exactly cancel to zero. We are then left with $7Y - 2X + 9$.
For this to tend to zero asymptotically we arrive at the condition,
\begin{equation}
-16X + 7c + 9 = 0
\end{equation}

Substituting the value of X from (29) and putting it in terms of $\Delta$ from (28) we have,
\begin{equation}
\frac{4}{\Delta} \frac{d \Delta}{d \tau} + 7c + 9 = 0
\end{equation}
Thus,
\begin{equation}
\frac{\Delta_{2}}{\Delta_{1}} = exp[(\frac{-9-7c}{4})(\tau_{2} - \tau_{1})]
\end{equation}
It is clearly seen from (42) that for c$>$ $- \frac{9}{7}$ $\Delta_{2} \rightarrow$ 0 for $\tau_{2}$$>>$$\tau_{1}$ and for c$<$ $-\frac{9}{7}$, $\Delta_{2} \rightarrow$ 0 for $\tau_{2}$$<<$ $\tau_{1}$.

Since the time scale can be arbitrarily adjusted , we will get a singularity in one of the cases mentioned above.

The conclusions arising from the fact that the volume element of the universe, $\Delta$ goes to zero in all possible cases and taken into consideration implies a singular solution to the generic Newtonian cosmological model with shear and rotation. The results show that in an expanding universe the outward force arising due to rotation can never be enough to overcome the combined force due to shear and gravitational pull. It may also be seen that the shear term cannot be taken equal to zero over the entire time scale if the expanding model of the universe is considered.

{\vskip 0.7in}
\centerline{\bf IV. CONCLUSION}
\medskip

Though singularities are unavoidable as shown by Hawking and Penrose (1975) in the general singularity theorem, nonetheless it is interesting to observe in these models how we arrive at singularities in the course of time. This result holds equally well for the Newtonian models with shear and rotation as they have been shown to be singular in the history or future of the universe. The centrifugal force arising due to the rotation of the universe can never adequately combat the gravitational and shearing force.

{\vskip 0.7in}
\centerline{\bf V.  ACKNOWLEDGEMENTS}
\medskip

I would like to thank Professor J.V. Narlikar, under whose guidance the work was undertaken, for discussions and constant support. I also thank all the members of the Inter University Centre for Astronomy and Astrophysics, Pune for their help  and the Jawaharlal Nehru Centre for Advanced Scientific Research, Bangalore for providing me the opportunity to undertake this work under the Summer Research Fellowship Programme,1998.

{\vskip 0.7in}
\newpage
\centerline{\bf VI.  REFERENCES.}
\medskip

1. DAVIDSON W. and  EVANS A.B., 1973, Newtonian Universes Expanding or Contracting with Shear and Rotation, International Journal for Theoritical Physics, Vol. 7, No. 5, 353-378.
{\vskip 0.1in}

2. G$\ddot{O}$DEL K., 1949, Review of Modern Physics,21,447.
{\vskip 0.1in}

3. G$\ddot{O}$DEL K., 1950, Rotating universes in General Relativity Theory in Proceedings 1950 International Congress of Math Volume- I , 175-181.
{\vskip 0.1in}

4. ELLIS G.F., 1967, Dynamics of Pressure-Free Matter in General Relativity,J. Mathematical Physics 8, 1171.
{\vskip 0.1in}

5. HAWKING S. and ELLIS G.F., 1975, Large Scale Structure of Space-Time, Cambridge University Press.
{\vskip 0.1in}

6. HECKMANN O. and SCH$\ddot{U}$CKING E., 1955, Zeitschrift f$\ddot{u}$r Astrophysik, 38, 95.
{\vskip 0.1in}

7. HECKMANN O. and SCH$\ddot{U}$CKING E., 1958, Solvay Conferences (Brussels).
{\vskip 0.1in}

8. McCREA W.H. and MILNE E.A., 1934, Quaterly Journal of Mathematics, 5, 73.
{\vskip 0.1in}

9. NARLIKAR J.V., 1960, Estratto da Rendiconti della Seuola Internazionale di Fisica - XX Corso, 222-227,2.
{\vskip 0.1in}

10. NARLIKAR J.V., 1963, Newtonian Universes with Shear and Rotation, Monthly Notices of the Royal Astronomical Society, Vol. 126, 203-208.
{\vskip 0.1in}

11. RAYCHAUDHURI A.K., 1955, Physical Review, 98, 1123.
{\vskip 0.1in}

\begin{figure}
\begin{center}
\mbox{\epsfig{figure=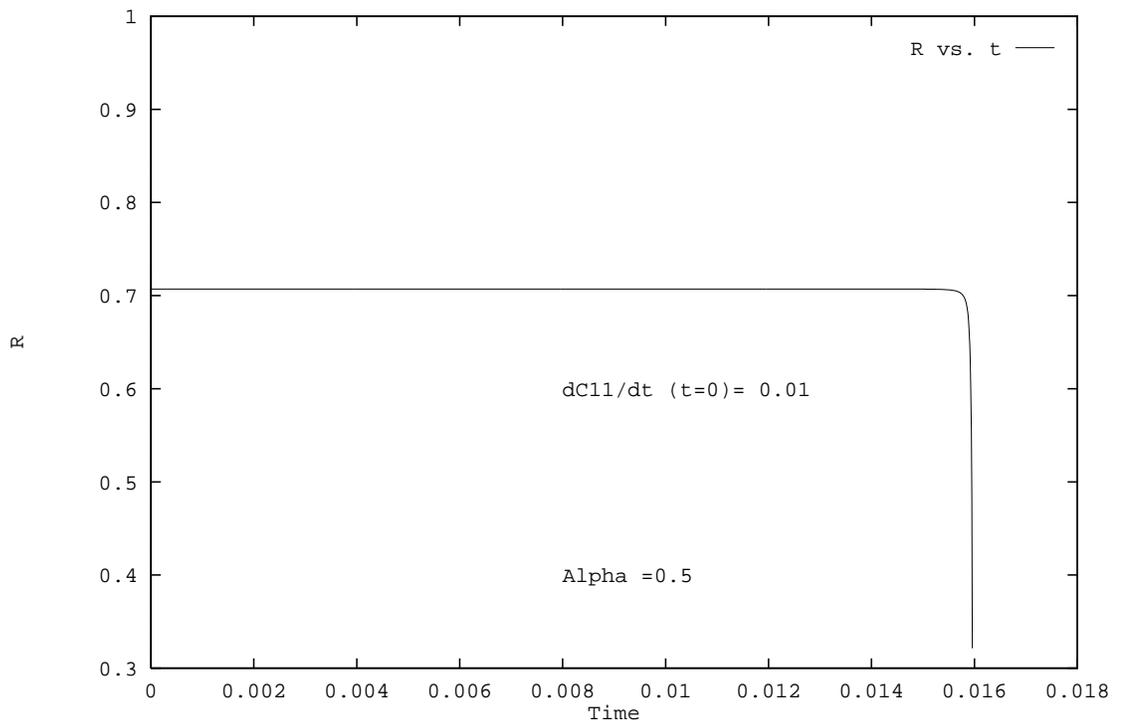,width=4.1in,angle=-90}}
\caption{{\bf Variation of $R$ with time with $\alpha=0.5$.} The figure shows the behaviour of $R$ with time for a positive $\alpha$ and reveals a behaviour  culminating in a collapse.}
\label{Figure 1}
\end{center}
\end{figure}

\begin{figure}
\begin{center}
\mbox{\epsfig{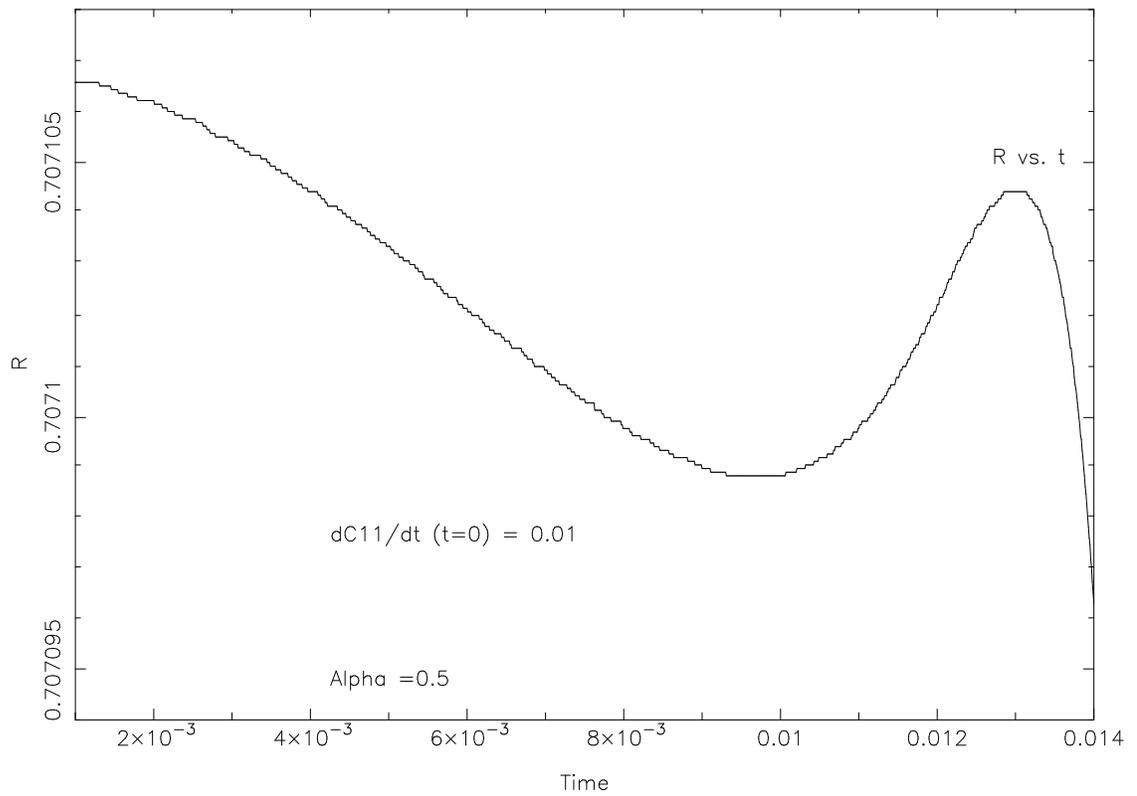}}
\caption{{\bf Variation of $R$ with time with $\alpha=0.5$.} The figure shows that during the initial time the behaviour of $R$ for a positive $\alpha$ reveals an oscillating pattern.}
\label{Figure 2}
\end{center}
\end{figure}

\begin{figure}
\begin{center}
\mbox{\epsfig{figure=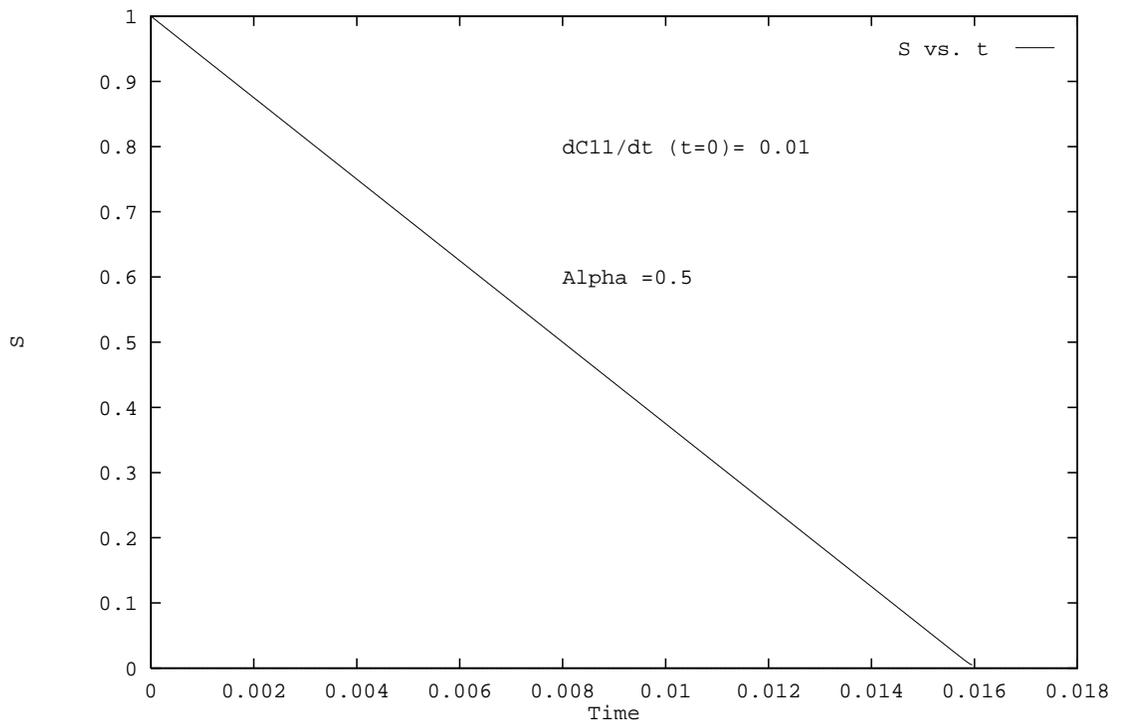,width=4.1in,angle=-90}}
\caption{{\bf Variation of $S$ with time with $\alpha=0.5$.}  The figure shows the behaviour of the coefficient of the $x^{3}$ dimension, for a positive $\alpha$, which is seen to go to zero linearly with time.}
\label{Figure 3}
\end{center}
\end{figure}

\begin{figure}
\begin{center}
\mbox{\epsfig{figure=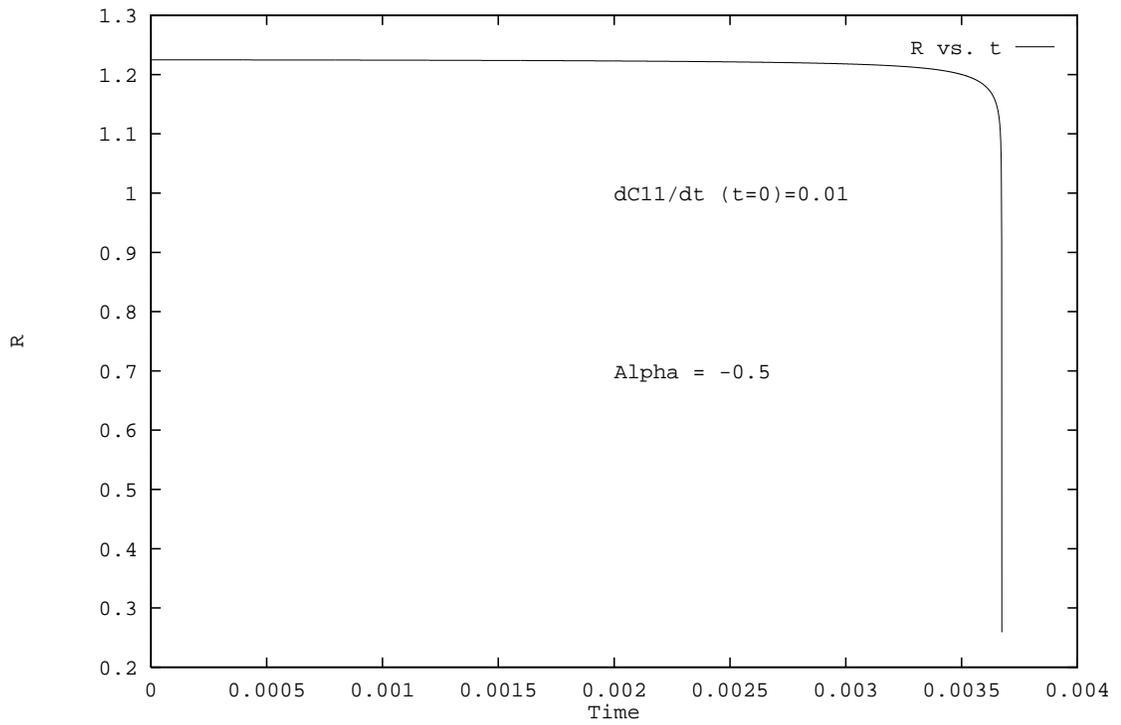,width=4.1in,angle=-90}}
\caption{{\bf Variation of $R$ with time with $\alpha =-0.5$.} The figure shows that for a negative $\alpha$, $R$ goes to zero monotonically without executing any oscillations about an order of magnitude faster than the case with a positive $\alpha$.}
\label{Figure 4}
\end{center}
\end{figure}

\begin{figure}
\begin{center}
\mbox{\epsfig{figure=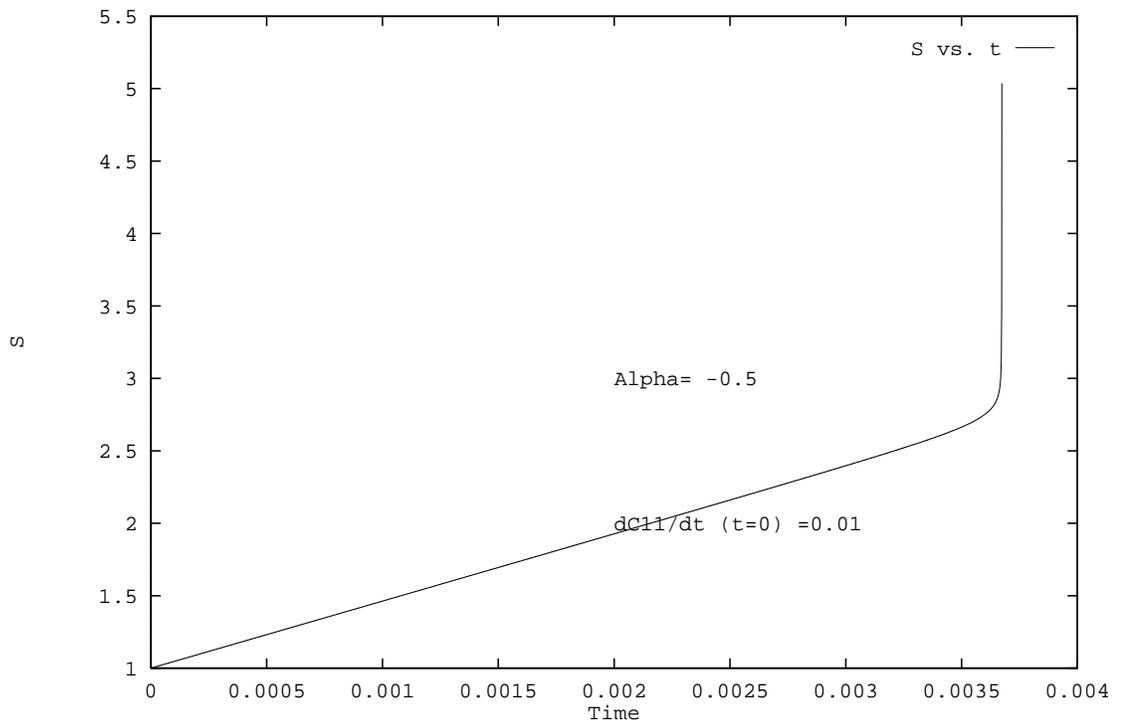,width=4.1in,angle=-90}}
\caption{{\bf Variation of $S$ with time with $\alpha =-0.5$.}  The figure shows that for a negative $\alpha$, the coefficient of the $x^{3}$ dimension  increases with time first linearly and then very rapidly.} 
\label{Figure 5}
\end{center}
\end{figure}

\begin{figure}
\begin{center}
\mbox{\epsfig{figure=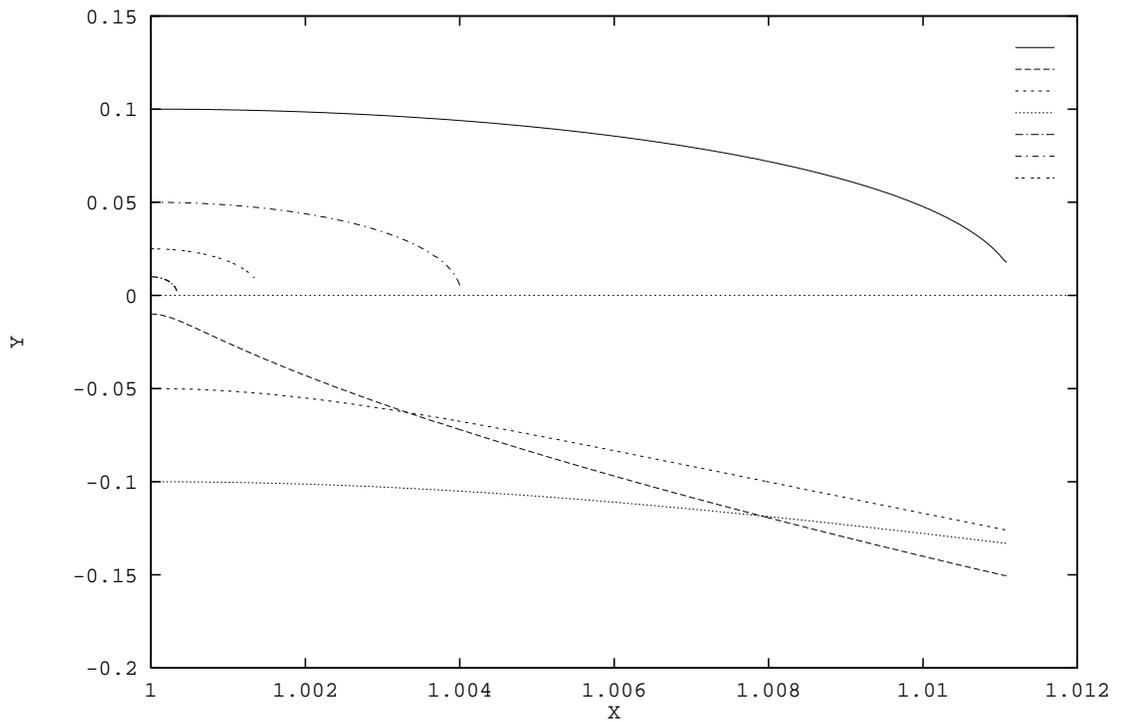,width=4.1in,angle=-90}}
\caption{{\bf Variation of X with Y for some different initial conditions.} The curves show that the curves for $Y>0$ can never be joined to the curves for $Y<0$ and that $Y$ eventually decreases with $X$ corresponding to universes where rotataion has been ineffective in preventing a singular state. All curves in the figure asymptotically go as $Y=-2X$.}
\label{Figure 6}
\end{center}
\end{figure}

\end{document}